\ifdefined\pdfoutput\pdfoutput=1\fi
\documentclass[11pt,letterpaper]{article}

\usepackage[letterpaper,textwidth=430pt,textheight=666pt,centering]{geometry}

\usepackage[T1]{fontenc}
\usepackage{times}
\usepackage{graphicx}
\usepackage{amsmath,amssymb}
\usepackage{booktabs}
\usepackage{array}
\usepackage{calc}
\usepackage{url}

\providecommand{\tightlist}{\setlength{\itemsep}{0pt}\setlength{\parskip}{0pt}}
\providecommand{\real}[1]{#1}
\providecommand{\PARstart}[2]{#1#2}
\title{EUReCA: End-to-End Unified CBCT Reconstruction Across Clinical Acquisitions}
\author{Jiening Zhu, Chengzhu Zhang, Jason Fan, LiCheng Kuo, Weixing Cai,\\ Xiuxiu He, Laura Cervino, Jean Moran, Xiang Li, Tianfang Li, and Yabo~Fu%
\thanks{This work was supported in part by the National Cancer Institute of the National Institutes of Health under Award P30 CA008748. (Corresponding author: Yabo Fu.)}%
\thanks{The authors are with the Department of Medical Physics, Memorial Sloan Kettering Cancer Center, New York, NY 10065 USA (e-mail: fuy3@mskcc.org).}%
}
\date{}

\begin{document}
\maketitle

\begin{abstract}
Cone-beam computed tomography (CBCT) is widely used for image-guided
radiotherapy, but reducing the number of projections to lower dose and
scan time makes reconstruction increasingly ill-posed. We present EUReCA
(End-to-end Unified REconstruction across Clinical Acquisitions), a
geometry-aware feed-forward model for variable-view clinical linac CBCT.
For a given acquisition trajectory, EUReCA accepts 8--100 projections
with a single set of weights, without view-count-specific retraining;
the same architecture and training recipe extend unchanged to other
clinical trajectories. It combines learned projection features with an
in-frame Feldkamp--Davis--Kress (FDK) reconstruction. The model encodes
each projection with a shared multiscale backbone, back-projects
features into multiresolution 3D grids using the acquisition geometry,
aggregates variable numbers of views at fixed computational cost, and
refines the volume with a global 3D transformer and multilevel decoder.
On simulated data, EUReCA achieved 29.9--33.2 dB PSNR with 10--100 views
and outperformed all applicable feed-forward baselines; it also exceeded
iterative optimization-based methods while running orders of magnitude
faster. On 15 measured patient CBCT scans evaluated against vendor
reconstructions, EUReCA achieved the best performance at every tested
view count. These results suggest that geometry-aware,
measurement-anchored feed-forward reconstruction can provide a practical
and flexible foundation for variable-view linac CBCT.
\end{abstract}

\noindent\textbf{Keywords:} cone-beam computed tomography, sparse-view
reconstruction, deep learning, image-guided radiotherapy.
\medskip

\section{Introduction}\label{introduction}

\PARstart{C}{one-beam} computed tomography (CBCT) provides three-dimensional patient
anatomy for image-guided radiotherapy. Conventional linac-based CBCT
acquires hundreds of projections over a full or partial gantry rotation.
Reducing the number of projections can lower imaging dose and shorten
acquisition time, but it also makes reconstruction increasingly
ill-posed: the problem shifts from well constrained under dense angular
sampling to severely underdetermined under sparse or incomplete
sampling, producing structured streaks and missing-wedge artifacts in
conventional reconstructions. A clinically useful reconstruction method
must therefore remain reliable across a wide range of sampling
conditions rather than at a single fixed operating point.

Measurement-driven methods reconstruct each volume primarily from its
acquired projections. Feldkamp--Davis--Kress (FDK) reconstruction
\cite{feldkamp1984fdk} is fast and deterministic but degrades
substantially under angular undersampling. Algebraic and regularized
iterative methods \cite{andersen1984sart, sidky2008asdpocs} combine
projection fidelity with hand-crafted priors, but require per-scan
optimization and careful parameter tuning. Neural fields and Gaussian
representations, including NAF \cite{zha2022naf}, IntraTomo
\cite{zang2021intratomo}, SAX-NeRF \cite{cai2024saxnerf}, and
R2-Gaussian \cite{zha2024r2gaussian}, instead fit a parameterized
volume to each acquisition, with the representation itself acting as an
implicit regularizer. These approaches remain closely anchored to the
measurements, but they typically require minutes to hours per volume,
and their constraints weaken as the number of views decreases.

Population-trained methods learn anatomical priors across datasets and
apply them through either iterative inference or direct feed-forward
reconstruction. Diffusion methods \cite{song2022inverse, chung2022mcg}
follow the first strategy, combining a learned prior with the acquired
projections through iterative posterior sampling. In principle, they can
represent multiple plausible reconstructions, although whether this
variability corresponds to clinically meaningful uncertainty remains
insufficiently validated. Their sampling procedures also require many
network evaluations, and memory constraints led early CBCT
implementations to slice-wise priors \cite{chung2023diffusionmbir};
recent samplers enforce consistency across slices and reconstruct full
volumes \cite{chung2024dds, song2024diffusionblend}, but per-volume
reconstruction still takes far longer than a single feed-forward pass.

The second strategy amortizes reconstruction in a population-trained
network, typically producing a volume without per-scan optimization.
DIF-Net \cite{lin2023difnet} estimates attenuation at arbitrary 3D
points from pixel-aligned multiview features, while geometry-aware
attenuation learning \cite{cui2023geoaware} explicitly back-projects 2D
features into a voxel grid. C2RV \cite{lin2024c2rv} extends this idea
with multiscale volumetric features and cross-view attention. Later
methods strengthen global 3D reasoning through more structured
representations: DIF-Gaussian \cite{lin2024difgaussian} models spatial
feature distributions with 3D Gaussians and adds test-time refinement;
X-LRM \cite{zhang2025xlrm} uses a multiview transformer to construct an
implicit X-triplane radiodensity field; X-GRM \cite{liu2025xgrm}
predicts voxel-anchored Gaussians that support both CT extraction and
differentiable projection rendering; and ILV \cite{lee2026ilv}
repeatedly updates an explicit latent volume with multiview evidence
before Gaussian decoding and volumetric refinement.

Despite this progress, existing feed-forward methods have not
demonstrated a single framework that spans the broad range of sampling
conditions encountered in linac CBCT. Many architectures are tied to a
fixed number of views or rely on attention mechanisms whose cost grows
rapidly with projection count and resolution. Under sparse sampling,
learned anatomical priors may also produce plausible but
measurement-unsupported structures. In addition, most methods are
developed and evaluated using projections simulated under simplified
acquisition trajectories rather than clinically realistic linac CBCT
geometries. Beyond reducing the number of views in a routine half-scan,
clinically constrained imaging may require qualitatively different
sparse acquisitions, such as two orthogonal views, four fixed gantry
views, or short arcs when scan time, dose, or clearance constraints
limit angular coverage. Comparisons across mismatched geometries, view
counts, datasets, and metric implementations provide limited evidence of
relative performance \cite{shi2026lowdose}, and transfer to measured
clinical projections remains insufficiently studied.

We address these limitations with EUReCA (End-to-end Unified
REconstruction across Clinical Acquisitions), a geometry-conditioned
feed-forward model for variable-view linac CBCT. EUReCA encodes
individual projections, back-projects multiscale features into
volumetric grids, and aggregates them into a fixed-size representation
for global 3D refinement. A multiscale decoder then combines learned
projection features with an FDK reconstruction derived from the same
measurements. Our contributions are:

\begin{enumerate}
\def\labelenumi{\arabic{enumi}.}
\tightlist
\item
  \textbf{Scalable geometry-aware reconstruction.} We introduce an
  end-to-end model that explicitly incorporates acquisition geometry
  while efficiently supporting variable view counts and trajectories.
\item
  \textbf{Adaptive FDK-reference fusion.} We incorporate a full 3D FDK
  reconstruction from the same projections as an analytic reference,
  using voxel-wise gates to adapt its contribution at each decoder
  scale.
\item
  \textbf{Expanded range and acquisition flexibility.} For a given
  trajectory, one set of weights spans ultra-sparse to well-sampled
  regimes (8--100 views). The same architecture and training recipe,
  trained once per trajectory, also cover two-view, four-view, and
  short-arc acquisitions without architectural changes.
\item
  \textbf{Controlled simulated and clinical evaluation.} We compare
  feasible methods under matched data, geometries, splits, and metrics,
  and evaluate them on measured patient projections against
  corresponding clinical reference reconstructions.
\end{enumerate}

Code, including the DRR projector (Apache-2.0), is available at
\url{https://github.com/Jiening666/EUReCA}, and the trained weights of
all reported models (CC BY-NC 4.0) at
\url{https://huggingface.co/jzhu35/EUReCA}. An interactive visualization
of the reconstructions is available at
\url{https://jiening666.github.io/cbct_recon/}. The internal prostate
cohort can be shared on request.

\section{Method}\label{method}

\subsection{Overview and problem
setup}\label{overview-and-problem-setup}

Let \(\{(p_i,\theta_i)\}\), \(i=1,\dots,N\), denote \(N\) cone-beam
projections, where view \(p_i\) is \(H\times W\) and is acquired at
gantry angle \(\theta_i\in[0,2\pi)\). The projections may be
nonuniformly distributed over a full or partial arc. Given a target CBCT
volume \(\mathbf{V}\) of size \(N_x\times N_y\times N_z\), we seek a
single feed-forward mapping
\[f_\phi:\{(p_i,\theta_i)\}_{i=1}^{N}\mapsto\mathbf{V}.\]The parameters
\(\phi\) may be shared across projection sequences with varying view
counts, angular distributions, and arc spans, enabling one model to
support the acquisition regimes represented during training without
view-count-specific retraining.

\begin{figure*}[!t]
\centering
\includegraphics[width=1.0\linewidth,keepaspectratio,alt={Method overview. (A) Geometry-aware lifting: each of the N projections is encoded by a shared pretrained backbone into a three-scale feature pyramid, and features at every scale are back-projected into 3D grids using the exact cone-beam geometry and pooled across views within angular sectors (Fig.\textasciitilde). (B) Coarse-to-fine 3D decoder: the coarse volume is refined by a global transformer, then decoded through MLP and \textbackslash times2 upsampling stages with multi-scale fusion to the final volume. (C) Analytic FDK reference: an in-frame FDK reconstruction of the same projections is encoded by a 3D CNN into the matching scale pyramid and merged into the decoder at every scale through gated fusion (\textbackslash otimes).}]{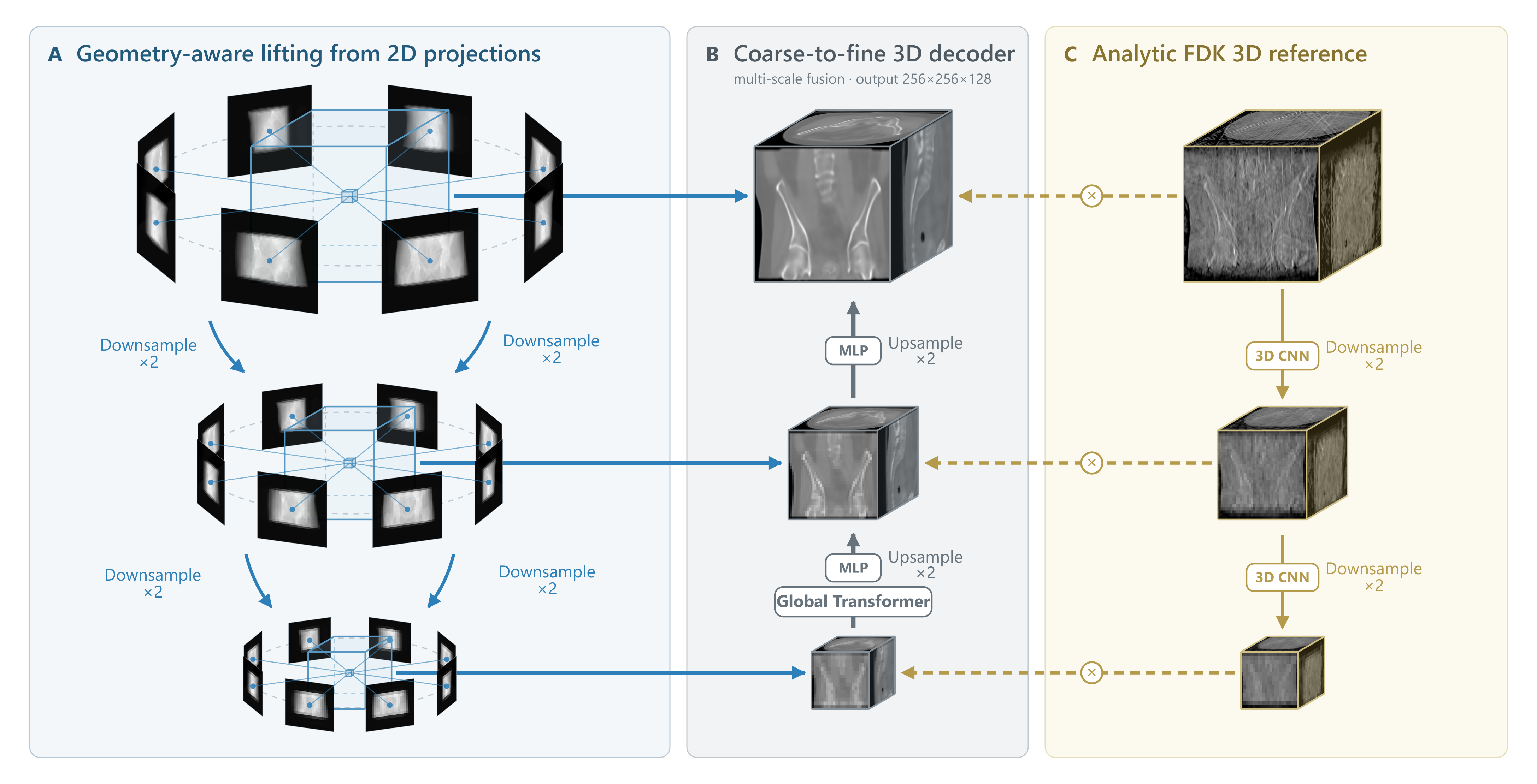}
\caption{\textbf{Method overview.} (A) \emph{Geometry-aware lifting:}
each of the \(N\) projections is encoded by a shared pretrained backbone
into a three-scale feature pyramid, and features at every scale are
back-projected into 3D grids using the exact cone-beam geometry and
pooled across views within angular sectors
(Fig.~\ref{fig:view_agg}). (B) \emph{Coarse-to-fine 3D
decoder:} the coarse volume is refined by a global transformer, then
decoded through MLP and \(\times2\) upsampling stages with multi-scale
fusion to the final volume. (C) \emph{Analytic FDK reference:} an
in-frame FDK reconstruction of the same projections is encoded by a 3D
CNN into the matching scale pyramid and merged into the decoder at every
scale through gated fusion (\(\otimes\)).}\label{fig:method_overview}
\end{figure*}

We realize the feed-forward mapping \(f_\phi\) as a geometry-aware
2D-to-3D network (Fig.~\ref{fig:method_overview}). A
shared multi-scale encoder extracts features from each projection.
Angular pooling then aggregates the variable number of views into
fixed-size 3D feature grids. A volumetric transformer captures global
anatomical context, while a U-Net-style decoder progressively restores
spatial detail. At each decoder scale, an in-frame FDK reconstruction is
incorporated as an analytic reference through gates that adapt to local
view coverage and overall view count. Together, these components enable
a single end-to-end model to reconstruct across view counts.

\subsection{Multi-scale projection
encoder}\label{multi-scale-projection-encoder}

Each projection is encoded into a three-level feature pyramid using an
ImageNet-pretrained ConvNeXt V2 Nano backbone \cite{woo2023convnextv2}.
The input layer is adapted to single-channel projections by averaging
its pretrained RGB weights. Features extracted at strides \(8\), \(16\),
and \(32\) form the fine, mid, and coarse levels, respectively;
\(1\times1\) convolutions followed by GroupNorm project them to \(C_f\),
\(C_m\), and \(C_c\) channels.

\subsection{Geometry-aware back-projection and cross-view
aggregation}\label{geometry-aware-back-projection-and-cross-view-aggregation}

\textbf{Back-projection.} For a target 3D grid at each scale, every
voxel is projected into every view using the exact cone-beam camera
geometry. The corresponding feature is bilinearly sampled, producing a
per-voxel stack of \(N\) view features (voxels projecting outside a
view's field of view are masked). Each view therefore contributes to a
voxel only through the ray that physically connects the source, the
voxel, and the detector.

\begin{figure*}[!t]
\centering
\includegraphics[width=1.0\linewidth,keepaspectratio,alt={Variable-view angular pooling over fixed sectors. (A) The N view features extracted by the shared pretrained 2D network are organized into K=36 fixed angular sectors by gantry angle: views within a sector are pooled to one sector feature, each carrying its angle position code, and unobserved sectors are represented explicitly (empty token), yielding an angle-encoded sector-feature sequence whose length is independent of N. (B) Per 3D location, sectors are aggregated scale-specifically: at the coarse scale a transformer self-attends across sectors through a class (CLS) token; at the mid and fine scales learned mixing weights drive a softmax-weighted aggregate. The learned aggregate is combined with mean and max statistics by a learned 1\textbackslash times1\textbackslash times1 fusion into the feature grid.}]{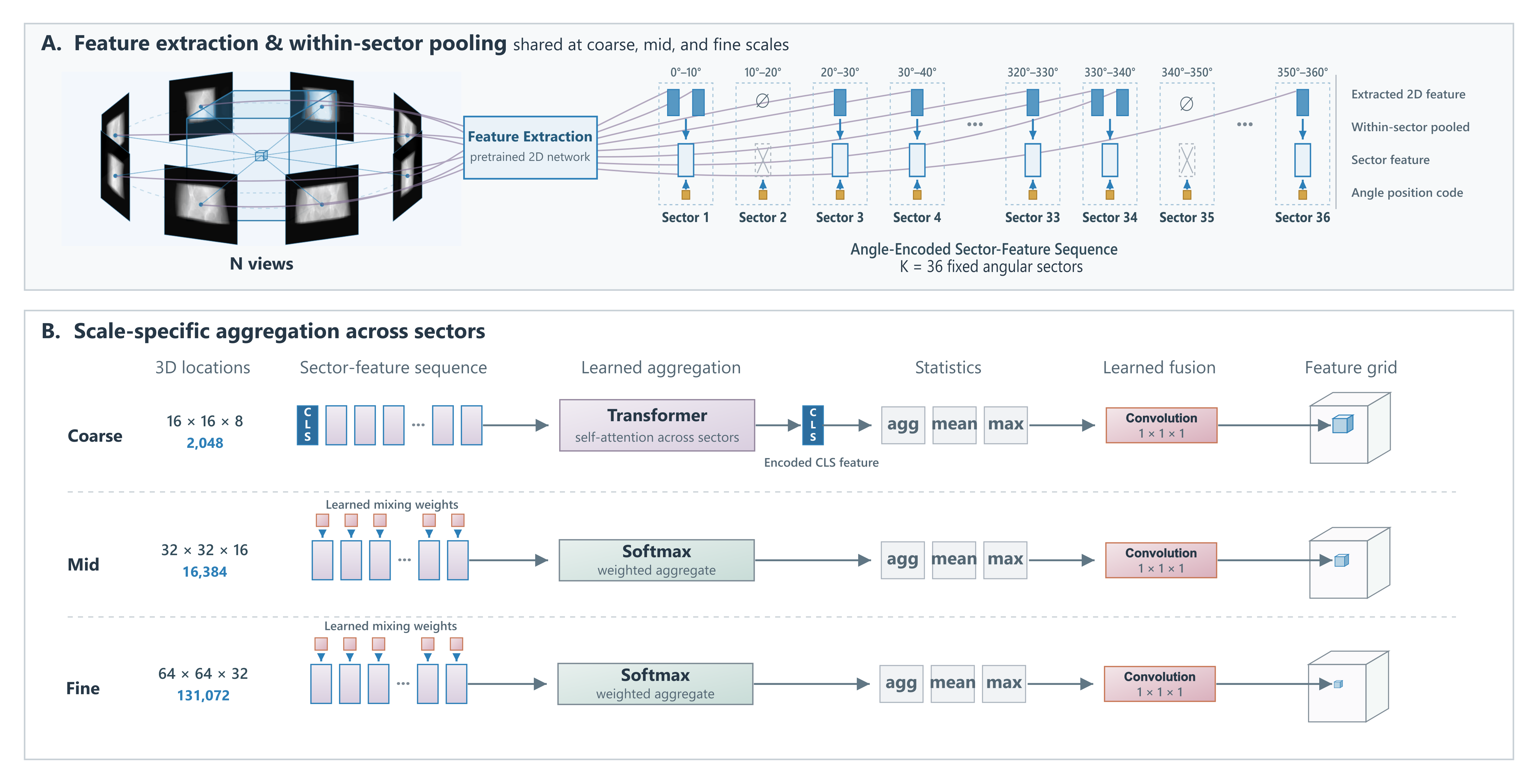}
\caption{\textbf{Variable-view angular pooling over fixed sectors.} (A)
The \(N\) view features extracted by the shared pretrained 2D network
are organized into \(K=36\) fixed angular sectors by gantry angle: views
within a sector are pooled to one sector feature, each carrying its
angle position code, and unobserved sectors are represented explicitly
(empty token), yielding an angle-encoded sector-feature sequence whose
length is independent of \(N\). (B) Per 3D location, sectors are
aggregated scale-specifically: at the coarse scale a transformer
self-attends across sectors through a class (CLS) token; at the mid and
fine scales learned mixing weights drive a softmax-weighted aggregate.
The learned aggregate is combined with mean and max statistics by a
learned \(1\times1\times1\) fusion into the feature
grid.}\label{fig:view_agg}
\end{figure*}

\textbf{Angular pooling.} Applying a transformer across all \(N\) views
costs \(O(N^2)\), while a single global softmax dilutes individual view
contributions as \(N\) grows. We instead pool in two learned steps over
a fixed partition of gantry angle into \(K=36\) angular sectors
(Fig.~\ref{fig:view_agg}A). Within a sector, a masked
softmax over learned per-view scores reduces the contributing views to
one feature per voxel, at \(O(N)\) cost. Across sectors, a second
learned aggregation combines the \(K\) sector features at every scale,
at a cost independent of \(N\).

Both steps act on back-projected per-voxel features, so pooling removes
angular redundancy at each voxel and leaves the spatial resolution of
the feature grids unchanged. The weights are learned per voxel from the
views' own features rather than being uniform, and back-projection
samples every view along its exact ray, so the angular detail the layer
can represent is not capped by the sector width; the FDK branch
described below receives every projection at its exact angle. The number
of sectors is a parameter of the layer rather than a property of the
method, and the three scales need not share one; all models reported
here use \(K=36\). The fixed-length sector sequence is what lets one set
of weights accept any number of views, while projection encoding,
back-projection, within-sector pooling and the FDK reconstruction all
scale linearly with \(N\), which is why reconstruction time still grows
with the view count (Table~II).

All three scales use the same sector-level geometric encoding. The
pooled feature for sector \(k\) is augmented with
\[[\sin\theta_k,\ \cos\theta_k,\ \log(1+n_k),\ c_k],\] where
\(\theta_k\) is the sector-center angle, \(n_k\) is its view count, and
\(c_k\in\{0,1\}\) indicates whether sector \(k\) or its conjugate sector
\((k+K/2)\bmod K\) is observed. This produces an angle-encoded sequence
of fixed length \(K\), independent of the number of input views.

\textbf{Cross-sector aggregation.} Features are aggregated across
angular sectors separately at each scale
(Fig.~\ref{fig:view_agg}B), with computation independent
of the number of input views. At the coarse scale, each sector token
additionally receives a 6-D Pl\"{u}cker ray encoding built from the voxel
position and the sector-center source direction. It is added to observed
and empty tokens alike, so the transformer also sees the geometry of the
directions that were never measured. An \(8\)-layer transformer then
processes a class token and the \(K\) sector tokens. Guided by angular
encodings and a conjugate-coverage cue, it learns interactions among
conjugate and neighboring directions without imposing a hard equivalence
between opposing rays. At the mid and fine scales, where the number of
spatial aggregation sites is much larger, learned mixing weights define
a lightweight softmax aggregation over occupied sectors. At each scale,
the resulting features are concatenated with the per-voxel mean and
maximum statistics. A learned \(1\times1\times1\) convolution then fuses
these inputs and reduces the concatenated channels back to the original
feature dimensionality, producing back-projected feature volumes
\(B^{(c)}\), \(B^{(m)}\), and \(B^{(f)}\) at 1/16, 1/8, and 1/4 of the
output resolution, respectively.

\subsection{Gated fusion of an in-frame analytic
reference}\label{gated-fusion-of-an-in-frame-analytic-reference}

Geometric back-projection maps 2D projection features into volumetric
grids but does not directly reconstruct the attenuation volume. We
therefore compute an FDK reconstruction \cite{feldkamp1984fdk} from the
same projections and fuse it into the decoder as an analytic reference.
It is inherently aligned, retains acquisition-specific structure under
sparse sampling, and becomes more reliable as angular coverage
increases. The FDK volume serves as an adaptive reference rather than an
exact data-consistency constraint.

\textbf{Analytic FDK reconstruction.} Our FDK implementation is
lightweight and contains no learned parameters. It uses fast Fourier
transform (FFT)-based filtered back-projection with a Hann-apodized ramp
filter. Angular-gap weighting handles nonuniform views, while Parker
weighting corrects redundancy in short scans. The volume is computed
once per forward pass without gradient tracking, requiring approximately
\(0.1\)~s. Under sparse or limited-angle sampling, FDK
exhibits severe streaking and missing-wedge shading but still preserves
gross anatomy, high-contrast boundaries, and spatial alignment. Its
reliability improves with angular coverage, allowing the gated decoder
to use it more strongly when supported by additional projections. Before
encoding, the volume is clipped at zero, normalized by its 99th
percentile, and bounded to \([0,1]\).

\textbf{Encoding the FDK volume and gated fusion.} A 3D ConvNeXt encodes
the FDK volume at three scales, producing feature volumes
\(\mathbf{F}^{(c)}\), \(\mathbf{F}^{(m)}\), and \(\mathbf{F}^{(f)}\). At
each scale \(\ell\), a voxel-wise gate fuses these features with the
back-projected projection features \(\mathbf{B}^{(\ell)}\). The gate is
a function of the learned features, FDK features, and local sampling
coverage:
\[\begin{aligned} g^{(\ell)} &= \sigma\!\left( \mathrm{Conv3D} \left[ \mathbf{B}^{(\ell)}, \mathbf{F}^{(\ell)}, \kappa^{(\ell)} \right] \right),\\ \mathbf{S}^{(\ell)} &= \mathbf{B}^{(\ell)} + g^{(\ell)}\odot \left(W^{(\ell)}\!\ast\mathbf{F}^{(\ell)}\right), \end{aligned}\]
where \(\kappa^{(\ell)}\) is the coverage map resampled to scale
\(\ell\), and \(W^{(\ell)}\) is a learned \(1\times1\times1\)
convolution. The coverage value combines the number and angular
diversity of views intersecting each voxel:
\[ \kappa = 1-\exp\!\left(-\frac{c}{\tau_{\mathrm{eff}}}\right), \qquad \tau_{\mathrm{eff}}=\frac{\tau_0}{\epsilon+s}, \]
where \(c\) is the in-field-of-view view count and \(s\) is the angular
spread, defined as one minus the mean resultant length of the
corresponding view angles. We use \(\tau_0=10\) and \(\epsilon=0.1\).
The FDK encoder is additionally conditioned on the global descriptor
\([\log N,s,\bar{\kappa}]\) by feature-wise linear modulation (FiLM)
\cite{perez2018film}, allowing feature extraction to adapt to view
count, angular spread, and mean coverage.

This design enables the contribution of the FDK reference to vary with
image content and local sampling support. We initialize \(W^{(\ell)}\)
near zero and bias the gate open, so the initial output remains close to
\(\mathbf{B}^{(\ell)}\) while the reference contribution is learned.
Ablation A3 evaluates the contribution of the FDK reference
(Sec.~\ref{sec:ablation}).

\subsection{Volume refinement and
decoding}\label{volume-refinement-and-decoding}

\textbf{Global volumetric interaction.} At the coarse scale, the fused
3D feature grid is flattened into a token sequence and processed by a
\(12\)-block transformer \(\mathrm{T}_{\mathrm{RoPE}}\). Each block
performs global self-attention and feed-forward refinement, while 3D
rotary positional embeddings (RoPE-3D) encode the spatial coordinates of
the grid tokens \cite{su2024rope}. Global token interaction captures
bilateral symmetry, organ-scale context, and relationships between
anatomically separated structures, promoting consistency between well-
and poorly sampled regions.

\textbf{Multi-scale decoding.} The decoder combines the fused projection
and FDK features \(\mathbf{S}^{(\ell)}\) with the decoder state from the
preceding coarser scale. Reconstruction proceeds coarse-to-fine:
\[\begin{aligned} \mathbf{d}^{(c)} &= \mathrm{T}_{\mathrm{RoPE}}\!\left(\mathbf{S}^{(c)}\right),\\ \mathbf{d}^{(m)} &= \phi^{(m)}\!\left( \left[ \mathrm{Up}\!\left(\mathbf{d}^{(c)}\right), \mathbf{S}^{(m)} \right] \right),\\ \mathbf{d}^{(f)} &= \phi^{(f)}\!\left( \left[ \mathrm{Up}\!\left(\mathbf{d}^{(m)}\right), \mathbf{S}^{(f)} \right] \right), \end{aligned} \]
where \(\cdot,\cdot\) denotes channel concatenation, \(\mathrm{Up}\) is
trilinear \(\times2\) upsampling, and \(\phi^{(\ell)}\) is a
convolutional block. Two additional skip-free upsampling blocks restore
full resolution:
\[\hat{\mathbf V} = W_o\, \psi_2\!\left( \mathrm{Up}\!\left[ \psi_1\!\left( \mathrm{Up}\!\left(\mathbf{d}^{(f)}\right) \right) \right] \right) \in \mathbb{R}^{N_x\times N_y\times N_z},\]
where \(\psi_1\) and \(\psi_2\) are convolutional blocks and \(W_o\) is
the final \(1\times1\times1\) projection. The output layer is linear.
Coarse global structure propagates through the decoder state, while the
multi-scale skips restore finer projection and FDK information.

\subsection{Training objective}\label{training-objective}

\textbf{Multi-scale gradient alignment (MAGA) loss.} Voxel-wise losses
encourage accurate intensities but may insufficiently penalize blurred
or displaced boundaries, which are particularly important for
soft-tissue interpretation. Inspired by multi-scale derivative
supervision for boundary recovery in monocular depth estimation
\cite{bochkovskii2025depthpro}, MAGA aligns the spatial gradients of
the prediction and target across multiple resolutions:
\[ \mathcal{L}_{\mathrm{MAGA}} = \frac{1}{S}\sum_{s=0}^{S-1} \left\lVert \mathbf{G}\,\pi_s(\hat{\mathbf V}) - \mathbf{G}\,\pi_s(\mathbf V) \right\rVert_1, \qquad S=3. \]
Here, \(\mathbf{G}(\cdot)=[\mathbf{G}_x,\mathbf{G}_y,\mathbf{G}_z]\)
denotes the 3D Scharr gradient operator. For each component, the
separable filter applies the derivative kernel \([-1,0,1]\) along the
corresponding axis and the smoothing kernel \(3,10,3\) along the other
two axes. The operator \(\pi_s(\cdot)=\mathrm{AvgPool}_{2^s}(\cdot)\)
downsamples the volume by a factor of \(2^s\). The finest scale
emphasizes precise boundary alignment, while the coarser scales capture
broader contrast transitions and shape discrepancies.

Many methods add a rendering loss that reprojects the predicted volume
and penalizes disagreement with the input views. With direct volume
supervision, we found that this loss provided negligible improvement
while more than doubling training time per step, so we omit it. The
ablation study evaluates both this choice and the contribution of
\(\mathcal{L}_{\mathrm{MAGA}}\)
(Sec.~\ref{sec:ablation}).

Training supervises the reconstructed volume directly, targeting the
error modes that matter clinically: absolute intensity accuracy,
structural fidelity, and sharp boundaries. The objective combines three
terms, \[
\mathcal{L} = \lambda_{\mathrm{SSIM}}\,(1-\mathrm{SSIM}_{\mathrm{tri}})
+ \lambda_{1}\,\lVert\hat{\mathbf V}-\mathbf V\rVert_1
+ \lambda_{\mathrm{MAGA}}\,\mathcal{L}_{\mathrm{MAGA}},
\] where \(\mathrm{SSIM}_{\mathrm{tri}}\) is the mean of slice-wise SSIM
over axial, coronal, and sagittal orientations (promoting structural
similarity), the \(L_1\) term anchors absolute intensities, and the
gradient-alignment term \(\mathcal{L}_{\mathrm{MAGA}}\) (above) sharpens
boundaries.

\section{Data and Experimental Setup}\label{data-and-experimental-setup}

\subsection{Training data}\label{training-data}

We assemble \(21{,}231\) CT volumes from seven public datasets and one
internal cohort spanning thoracic, abdominal, and pelvic anatomy: MELA
\cite{mela2022}, LUNA16 \cite{setio2017luna16}, AMOS22
\cite{ji2022amos}, RSNA-2023 \cite{rsna2023trauma}, AbdomenAtlas
\cite{li2024abdomenatlas, bassi2025radgpt}, TotalSegmentator
\cite{wasserthal2023totalseg}, RibFrac \cite{jin2020ribfrac}, and an
internal prostate cohort. To our knowledge, this is among the largest
datasets assembled for feed-forward-based CBCT reconstruction.

We retain only high-quality CT volumes with sufficient longitudinal
coverage, requiring voxel spacing below \(5\)~mm along
every axis and a craniocaudal extent greater than
\(300\)~mm. These criteria exclude low-resolution scans
unsuitable for high-fidelity digitally reconstructed radiograph (DRR)
generation and short scans in which simulated rays would intersect
artificial superior or inferior boundaries of the segmented body,
producing nonphysical projections. After filtering, \(12{,}255\) volumes
remain (Table~I).

\begin{table}[!t]
\scriptsize\setlength{\tabcolsep}{3pt}
\noindent \textbf{Table I. CT sources before and after filtering, with the
case-level split.}
\par\smallskip
\centering
\begin{tabular}{@{}lllllll@{}}
\toprule\noalign{}
Dataset
 & 
Region
 & 
Original
 & 
Filtered
 & 
Train
 & 
Val
 & 
Test
 \\
\midrule\noalign{}
MELA & Thorax (mediastinum) & 1,100 & 916 & 731 & 94 & 91 \\
LUNA16 & Thorax (lung) & 880 & 611 & 489 & 60 & 62 \\
RibFrac & Thorax (ribs) & 660 & 618 & 503 & 60 & 55 \\
AMOS22 & Abdomen & 2,383 & 796 & 654 & 75 & 67 \\
RSNA-2023 & Abdomen (trauma) & 4,138 & 2,462 & 1,922 & 259 & 281 \\
AbdomenAtlas & Abdomen & 9,262 & 5,998 & 4,828 & 586 & 584 \\
TotalSegmentator & Whole body & 1,228 & 756 & 603 & 75 & 78 \\
Prostate (internal) & Pelvis & 1,580 & 98 & 74 & 16 & 8 \\
\textbf{Total} & --- & \textbf{21,231} & \textbf{12,255} &
\textbf{9,804} & \textbf{1,225} & \textbf{1,226} \\
\bottomrule\noalign{}
\end{tabular}
\end{table}

\subsection{Acquisition geometry and simulated
projections}\label{acquisition-geometry-and-simulated-projections}

\begin{figure}[!t]
\centering
\includegraphics[width=1.0\linewidth,keepaspectratio,alt={Halcyon acquisition geometry. (a) The 211\ensuremath{^\circ} half-scan arc (gantry -126\ensuremath{^\circ} to 85\ensuremath{^\circ}), with the arc start free to rotate on the 360\ensuremath{^\circ} circle and colored markers illustrating independent sparse view draws; (b) cone-beam projection geometry with source-to-axis \textbackslash mathrm\{DSO\}=1000~mm, source-to-detector \textbackslash mathrm\{DSD\}=1540~mm, a 256\textbackslash times256\textbackslash times128 reconstruction volume at 2.0~mm isotropic voxels, and a 1.12~mm detector pixel pitch.}]{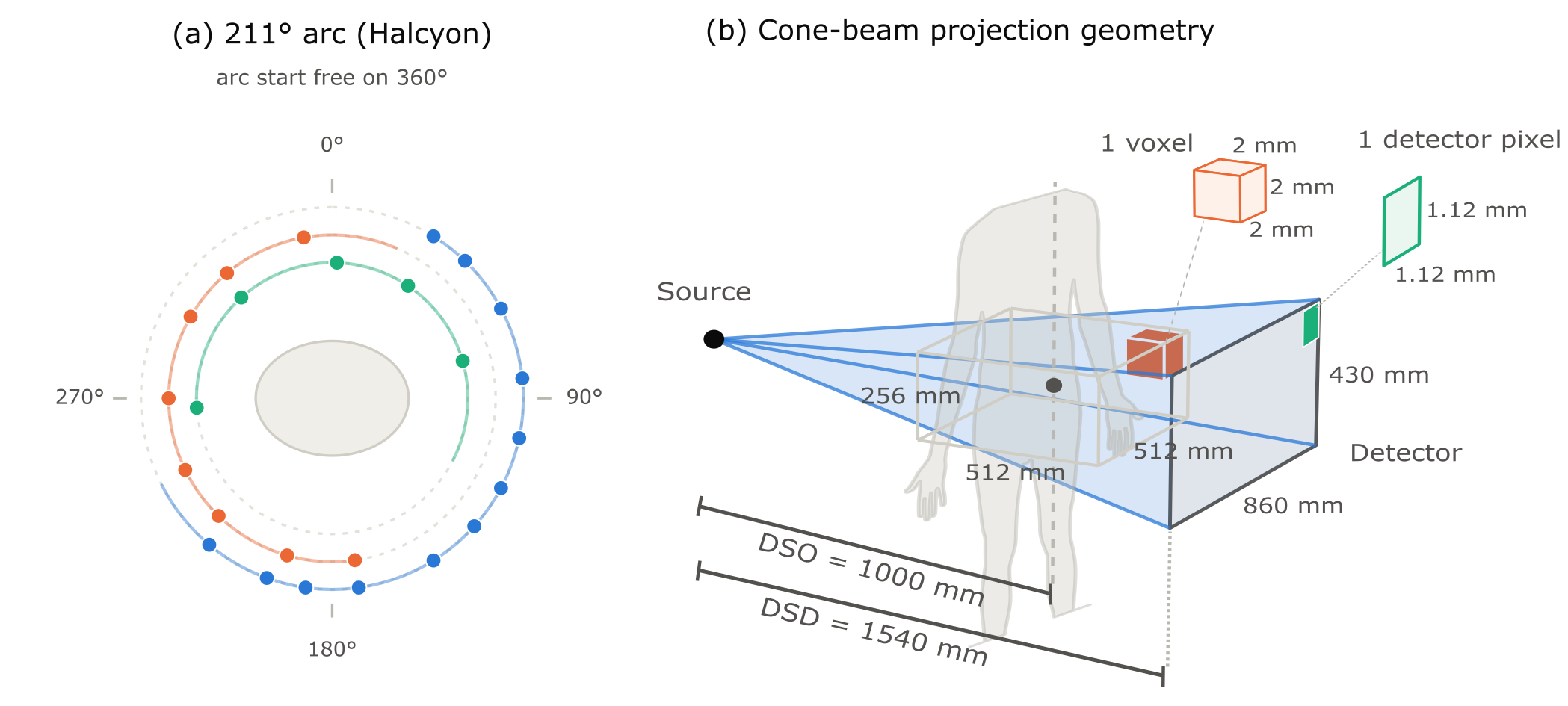}
\caption{\textbf{Halcyon acquisition geometry.} (a) The \(211\ensuremath{^\circ}\)
half-scan arc (gantry \(-126\ensuremath{^\circ}\) to \(85\ensuremath{^\circ}\)), with the arc start free to
rotate on the \(360\ensuremath{^\circ}\) circle and colored markers illustrating
independent sparse view draws; (b) cone-beam projection geometry with
source-to-axis \(\mathrm{DSO}=1000\)~mm,
source-to-detector \(\mathrm{DSD}=1540\)~mm, a
\(256\times256\times128\) reconstruction volume at
\(2.0\)~mm isotropic voxels, and a
\(1.12\)~mm detector pixel pitch.}\label{fig:geo_hal}
\end{figure}

The simulation geometry follows a clinical Varian Halcyon HyperSight
on-board imager rather than an idealized setup
(Fig.~\ref{fig:geo_hal}), reproducing its source-to-axis
distance, cone angle, magnification, and half-scan trajectory.

On this geometry we generate DRRs with a custom CUDA-accelerated
cone-beam projector that integrates attenuation along each
source-to-detector ray through the HU-to-attenuation-mapped volume,
rendering a \(384\times768\) detector of square \(1.12\)-mm pixels.
Square pixels are our one simplification of the scanner, whose native
detector bins anisotropically at \(1.12\times0.28\)~mm
over the same physical panel: the coarse axis is unchanged and only the
finer one resampled, giving an isotropic lattice that avoids
axis-dependent resolution. Rays are integrated through each CT at its
original resolution rather than through the resampled reconstruction
target, preserving projection fidelity. For each volume we render a
dense set of views over the full \(360^\circ\) circle, from which every
sparse acquisition is later drawn.

To model patient-positioning variability, projections are precomputed
for three poses: the original pose and two rigidly augmented poses. Each
augmentation applies rotations of up to \(\pm5^\circ\) about each axis
and translations within the margin between the scan extent and the
reconstruction field of view. The reconstruction isocenter remains fixed
at the volume center, but the transformations emulate changes in patient
position relative to the isocenter.

\subsection{Acquisition trajectories by
sampling}\label{acquisition-trajectories-by-sampling}

The primary acquisition is the routine patient-setup protocol, a
clinical Halcyon \(211^\circ\) half-scan sampled with angularly
stratified randomization. The arc may begin anywhere on the
\(360^\circ\) circle rather than at a fixed angle
(Fig.~\ref{fig:geo_hal}a). This protocol is used for the
main comparison, real-scan evaluation, and ablation studies. During
training, each sample draws a new acquisition from the precomputed dense
projection set: a random arc start, a view count sampled log-uniformly
from \([8,100]\), a new subset of angles within the arc, and one of
three precomputed poses.

\begin{figure*}[!t]
\centering
\includegraphics[width=1.0\linewidth,keepaspectratio,alt={Four additional clinical acquisitions --- two orthogonal views, four fixed views, and 90\ensuremath{^\circ}/120\ensuremath{^\circ} short arcs; markers show independent draws with the arc start free on the 360\ensuremath{^\circ} circle. Each is trained with the same architecture and recipe; only the view-selection trajectory changes at data loading.}]{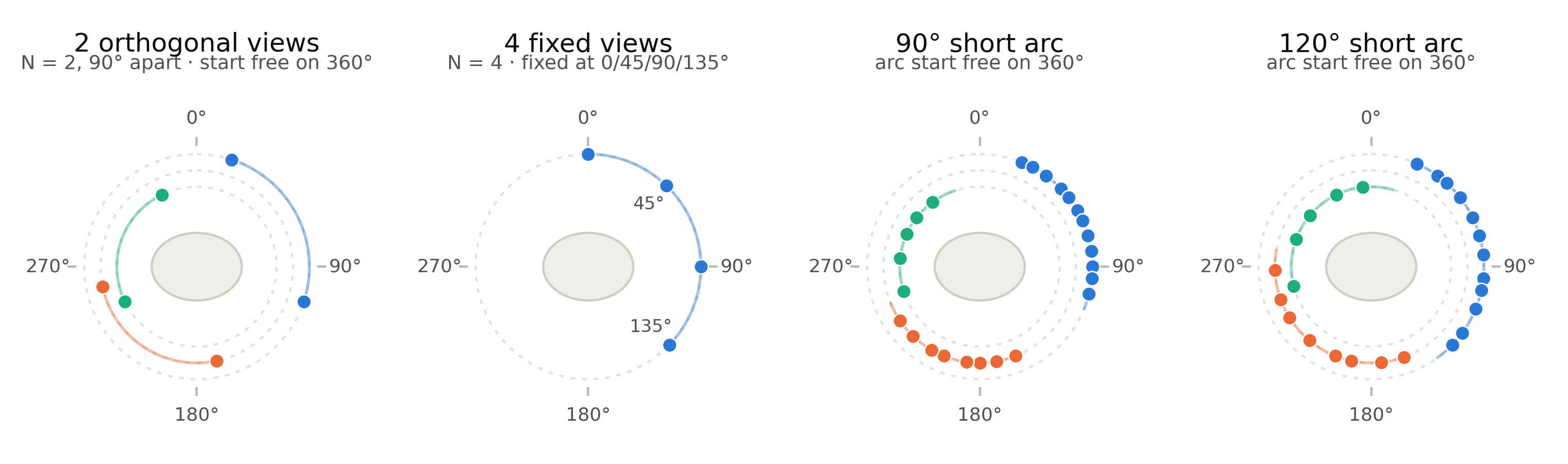}
\caption{\textbf{Four additional clinical acquisitions} --- two
orthogonal views, four fixed views, and \(90\ensuremath{^\circ}\)/\(120\ensuremath{^\circ}\) short arcs;
markers show independent draws with the arc start free on the \(360\ensuremath{^\circ}\)
circle. Each is trained with the same architecture and recipe; only the
view-selection trajectory changes at data loading.}\label{fig:acq_geom}
\end{figure*}

We additionally evaluate \(90^\circ\) and \(120^\circ\) short arcs,
four-view fixed-angle acquisitions, and two orthogonal views
(Fig.~\ref{fig:acq_geom}), representing fast or
clearance-limited imaging and fixed-angle setup verification. Uniform
and random angular sampling are also included as reference patterns.
Across all settings, only the view-selection trajectory changes; the
architecture, loss function, and training procedure remain unchanged.

To reproduce clinical beam collimation, we also apply projection-domain
blade blocking during training. With probability \(0.2\), up to \(7\%\)
of the detector is masked along its upper and lower edges. The real-scan
evaluation includes the same effect because the vendor pipeline masks
the collimator blades before detector binning.

\subsection{Data splits}\label{data-splits}

All models and baselines share a single case-level \(80\):\(10\):\(10\)
partition of the \(12{,}255\) cases (\(9{,}804\) training, \(1{,}225\)
validation, \(1{,}226\) test; Table~I), drawn once and
written to disk, so it is identical across every run and comparator.
Case-level splitting keeps all scans of a study together; the reserved
test set is used only for final evaluation, never for training or model
selection. Evaluation always uses the original pose for validation and
test.

\subsection{Real clinical projections}\label{real-clinical-projections}

We evaluate the DRR-trained model on measured clinical projections from
\(15\) patients scanned with a Varian Halcyon HyperSight on-board
imager. The CBCT acquisitions span head-and-neck (tonsil, tongue,
thyroid, and general H\&N), thoracic (esophagus), and pelvic (prostate
bed, bladder, rectum, uterus, and pubis) sites, all acquired using the
routine patient-setup protocol described above. Because these anatomical
sites are underrepresented in the training corpus, the evaluation tests
both DRR-to-kV transfer and anatomical generalization.

The raw projections undergo the vendor's standard preprocessing,
including detector-defect, scatter, beam-hardening, and spectrum
corrections followed by negative-log conversion. Four adjacent detector
pixels are then binned along the finer axis to match the model's square
\(1.12\)-mm input grid. Because no independent ground-truth CT is
available, reconstructions are evaluated against the vendor's clinical
iTools reconstruction, which uses all \(335\)--\(407\) projections from
each acquisition, whereas the evaluated methods receive only
\(10\)--\(100\). The vendor reconstruction is treated as a clinical
reference rather than ground truth, and scores are computed over the
intersect physically valid region of the reconstruction grid. Thus, the
reported metrics quantify agreement with the clinical standard.

\subsection{Baselines and harmonized
protocol}\label{baselines-and-harmonized-protocol}

We compare EUReCA with eight methods using harmonized data, acquisition
geometry, splits, and evaluation code wherever their formulations
permit. The three feed-forward (FF) baselines---DIF-Net
\cite{lin2023difnet}, DIF-Gaussian \cite{lin2024difgaussian}, and ILV
\cite{lee2026ilv}---are adapted to our reconstruction grid, HU window,
dataset filtering, patient-setup sampling, augmentations, and data
split, and are evaluated on the full test set. We also reimplemented the
geometry-aware attenuation model \cite{cui2023geoaware} and validated
its projection geometry. However, its released single-GPU pipeline did
not converge at our data scale within the fixed compute budget; we
therefore exclude it rather than report a compute-limited result. X-LRM
\cite{zhang2025xlrm} and X-GRM \cite{liu2025xgrm} are not compared
because neither has a public implementation; both author repositories
were empty or stated that code was forthcoming when checked on
2026-10-26.

View-count training cannot be fully harmonized because none of the
learning-based baselines was designed for variable-view training. We
therefore train separate models at \(10\), \(20\), \(50\), and \(100\)
views wherever feasible. Comparisons end at the highest count supported
by each architecture because fixed-view components or memory
requirements preclude some larger settings. In contrast, EUReCA uses a
single set of weights trained with view counts sampled log-uniformly
from \([8,100]\) and requires no retraining at evaluation. Unless
explicitly labeled a fixed-count \emph{specialist}, ``ours'' refers to
this single variable-view model. To measure any advantage from
count-specific training, we also train EUReCA specialists at the same
four view counts using an otherwise identical recipe and report the
comparison in ablation Study 3.

The remaining five are measurement-driven (MD): they require no
dataset-level training and operate independently on each scan: analytic
FDK, using the same configuration as our internal reference, and four
optimization-based methods---R2-Gaussian \cite{zha2024r2gaussian}, NAF
\cite{zha2022naf}, IntraTomo \cite{zang2021intratomo}, and SAX-NeRF
\cite{cai2024saxnerf}. Because these optimization-based methods require
minutes to hours per volume, we evaluate them on a fixed, stratified
subset of \(20\) test scans using their published optimization budgets.

\subsection{Evaluation metrics}\label{evaluation-metrics}

Reconstructions are scored by volumetric PSNR and tri-planar SSIM
(slice-wise SSIM averaged over axial, coronal, and sagittal
orientations; data range \(1\)), which extends a standard 2D SSIM
implementation to volumes by evaluating it on all three orthogonal slice
orientations; pixel-wise metrics can reward over-smoothing
\cite{lin2025care}, a limitation we return to in the Discussion.
Because tri-planar SSIM is also a term of our training loss, it may
favor our model; PSNR, which no method optimizes directly, is therefore
the primary metric. Every method writes floating-point
prediction/ground-truth volumes scored by one shared metric
implementation.

\subsection{Implementation details}\label{implementation-details}

The model is trained with HuggingFace Accelerate (DDP, \texttt{bf16}) on
two NVIDIA A100 GPUs at per-GPU batch size one, using AdamW
(\(\beta=(0.9,0.95)\)), base learning rate \(2\times10^{-4}\)
cosine-annealed over \(50\) epochs, gradient clipping at \(1.0\), and a
\(0.1\times\) multiplier on the pretrained backbone. Each epoch draws
five sparse-view samples per training volume
(\(\approx2.4\)~million over the run), each independently
sampling its view count log-uniformly in \([8,100]\) and its angles from
the experiment's trajectory. Peak memory at large view counts is bounded
by gradient-checkpointed, view- and voxel-chunked encoding and
aggregation (encoder view chunk \(16\)). The reconstruction target is
each CT resampled to \(256\times256\times128\) at
\(2.0\)~mm isotropic spacing (a
\(512\times512\times256\)~mm field, following the
reconstruction setting of a clinical Halcyon CBCT protocol), windowed to
\([-1000,1000]\)~HU and normalized to \([0,1]\), with
back-projected feature grids at \(1/16\), \(1/8\), and \(1/4\) of that
resolution; feature-pyramid channel widths are
\((C_f,C_m,C_c)=(128,256,768)\) at the fine, mid, and coarse scales; and
the internal FDK and DRR operators share the acquisition geometry of the
simulation. Loss weights follow the ablation-justified baseline:
\((\lambda_{\mathrm{SSIM}},\lambda_{1},\lambda_{\mathrm{MAGA}})=(50,50,1)\)
and the projection-consistency render loss off. The fixed-count
specialists of Study 3 reuse this configuration with the view count held
fixed; the four-view experiment instead uses four fixed gantry angles
(\(0\ensuremath{^\circ}/45\ensuremath{^\circ}/90\ensuremath{^\circ}/135\ensuremath{^\circ}\), snapped to the projection grid) spanning a
\(135\ensuremath{^\circ}\) arc.

\section{Results}\label{results}

We evaluate our framework in two stages: first on simulated projections
and then on real clinical projections. Using synthetic DRR data, we
benchmark reconstruction quality and speed against feed-forward (FF) and
measurement-driven (MD) baselines under the routine patient-setup
protocol, for which the full suite of comparators is trained. We then
use the same synthetic DRR data to demonstrate the framework across
additional clinically relevant acquisition settings. Finally, we
quantitatively evaluate performance on measured clinical projections
acquired from real patients using the routine patient-setup protocol.

\subsection{Comparison with state of the art on the routine
patient-setup
protocol}\label{comparison-with-state-of-the-art-on-the-routine-patient-setup-protocol}

Following our reporting protocol, each method is trained and tested at
matched view counts (its published operating regime); a cell is marked
``n/a'' where an architecture cannot reach that count.

\begin{table*}[!t]
\footnotesize
\noindent \textbf{Table II. Quality and speed vs.~view count (PSNR dB / SSIM;
routine patient-setup protocol). Times are end-to-end wall-clock per
volume on a single GPU, including projection preprocessing, the internal
FDK reconstruction, network inference, and host--device transfer.}
\par\smallskip
\centering
\begin{tabular}{@{}
  >{\raggedright\arraybackslash}p{(\linewidth - 12\tabcolsep) * \real{0.1322}}
  >{\raggedright\arraybackslash}p{(\linewidth - 12\tabcolsep) * \real{0.0441}}
  >{\raggedright\arraybackslash}p{(\linewidth - 12\tabcolsep) * \real{0.1394}}
  >{\raggedright\arraybackslash}p{(\linewidth - 12\tabcolsep) * \real{0.1394}}
  >{\raggedright\arraybackslash}p{(\linewidth - 12\tabcolsep) * \real{0.1394}}
  >{\raggedright\arraybackslash}p{(\linewidth - 12\tabcolsep) * \real{0.1394}}
  >{\raggedright\arraybackslash}p{(\linewidth - 12\tabcolsep) * \real{0.2661}}@{}}
\toprule\noalign{}
Method
 & 
Type
 & 
10 views
 & 
20 views
 & 
50 views
 & 
100 views
 & 
Time 10/20/50/100v
 \\
\midrule\noalign{}
\textbf{Ours} & FF & \textbf{29.93/0.908} & \textbf{31.52/0.928} &
\textbf{32.77/0.944} & \textbf{33.22/0.952} &
\textbf{0.32/0.41/0.71/1.33s} \\
ILV & FF & 29.42/0.875 & 30.17/0.891 & n/a & n/a & 0.38/0.50s/OOM/OOM \\
DIF-Net & FF & 23.91/0.724 & 25.19/0.746 & 26.71/0.786 & n/a &
1.73/5.00/14.9s/--- \\
DIF-Gaussian & FF & 24.39/0.741 & 23.30/0.611 & 22.76/0.560 & n/a &
2.06/5.64/16.9s/--- \\
FDK & MD & 15.34/0.252 & 17.76/0.338 & 21.13/0.579 & 22.45/0.740 &
0.11/0.10/0.08/0.17s \\
R\ensuremath{^2}-Gaussian & MD & 20.55/0.550 & 23.06/0.661 & 24.77/0.768 & 24.86/0.786
& 4.6/4.5/4.6/15.0min \\
IntraTomo & MD & 20.53/0.572 & 23.71/0.751 & 25.25/0.822 & 25.78/0.845 &
2.9/5.1/12.1/17.4min \\
NAF & MD & 21.76/0.595 & 23.89/0.771 & 25.41/0.828 & 25.84/0.845 &
6.5/11.9/27.3/38.3min \\
SAX-NeRF & MD & 22.41/0.752 & 24.41/0.805 & 25.93/0.845 & 27.92/0.887 &
0.45/0.49/1.19/3.56h \\
\bottomrule\noalign{}
\end{tabular}
\end{table*}

Our model achieved the highest PSNR and SSIM at every view count. Among
feed-forward baselines, ILV trails by \(0.51\)~dB at
\(10\) views and \(1.35\)~dB at \(20\) views, requires a
separate model for each view count, and runs out of memory beyond \(20\)
views. At \(50\) views, our model outperformed the best remaining
feed-forward baseline by \(6.06\)~dB; at \(100\) views,
no feed-forward baseline supports the setting. Our model also exceeds
the strongest measurement-driven method by more than
\(7\)~dB at \(10\) views and by at least
\(3.6\)~dB at \(20\)--\(100\) views, while being orders
of magnitude faster. DIF-Gaussian exhibits decreasing accuracy as the
view count increases, with PSNR falling from \(24.39\) to
\(22.76\)~dB between \(10\) and \(50\) views. One
possible explanation is that the method was validated only at up to
\(10\) views in the original publication, while its max-pooling
aggregation retains only the dominant response at each location,
potentially limiting the contribution of additional views and reducing
stability beyond the published operating regime. Finally, our model
reconstructs a volume in \(0.3\)--\(1.3\)~s, making it
two to four orders of magnitude faster than the per-scan optimizers. ILV
provides comparable speed only at low view counts, whereas the DIF
models are \(5\)--\(21\times\) slower and scale less favorably with
additional views.

\begin{figure*}[!t]
\centering
\includegraphics[width=1.0\linewidth,keepaspectratio,alt={Qualitative comparison at 10 views (thoracic case, axial/coronal/sagittal). Top: feed-forward methods --- DIF-Net, DIF-Gaussian, ILV, ours, ground truth. Bottom: measurement-driven methods --- FDK, R\ensuremath{^2}-Gaussian, IntraTomo, NAF, SAX-NeRF, ours, ground truth.}]{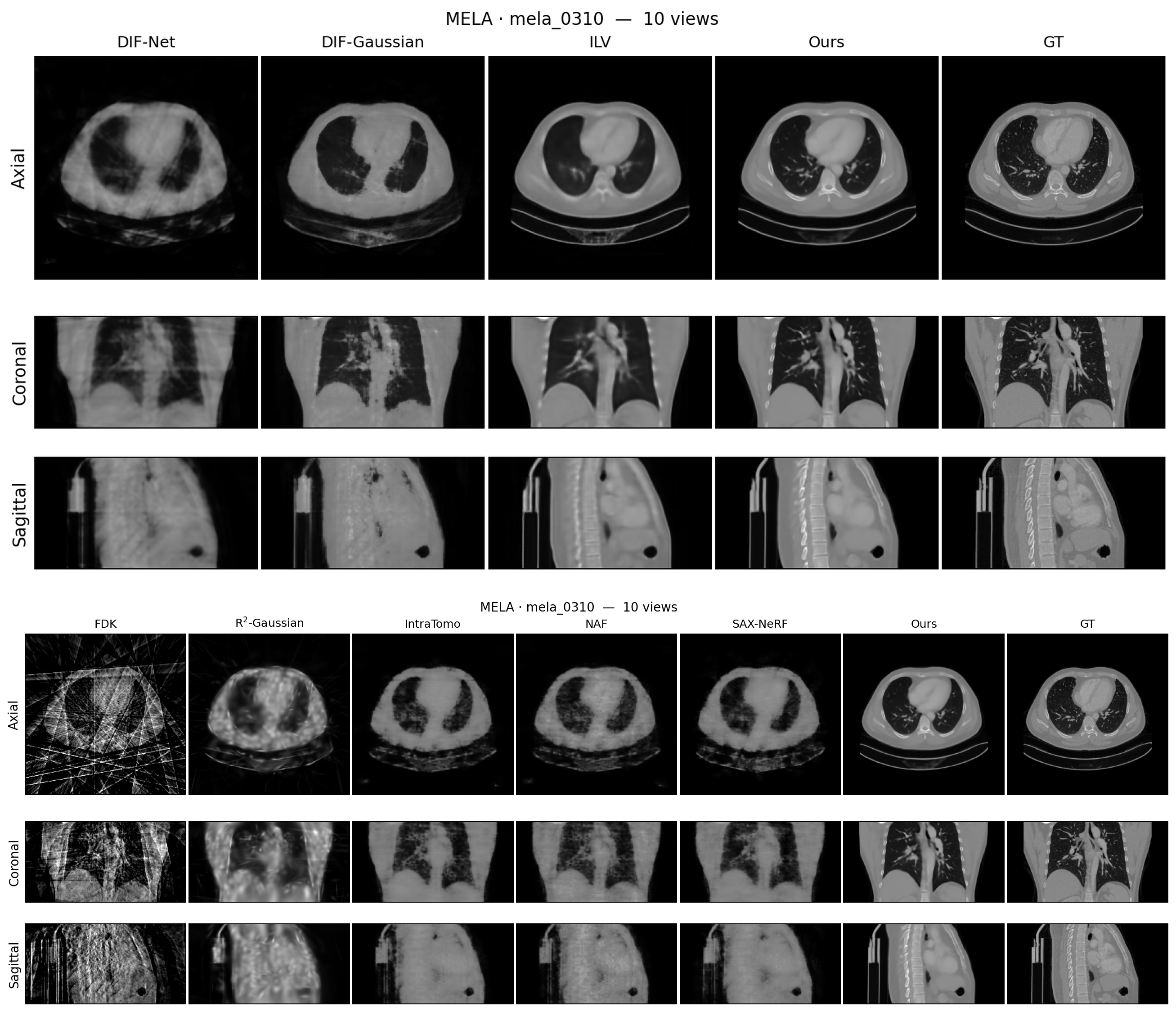}
\caption{\textbf{Qualitative comparison at 10 views} (thoracic case,
axial/coronal/sagittal). \emph{Top:} feed-forward methods --- DIF-Net,
DIF-Gaussian, ILV, ours, ground truth. \emph{Bottom:} measurement-driven
methods --- FDK, R\ensuremath{^2}-Gaussian, IntraTomo, NAF, SAX-NeRF, ours, ground
truth.}\label{fig:qual}
\end{figure*}

Figure~\ref{fig:qual} presents axial, coronal, and
sagittal reconstructions of a representative thoracic case at \(10\)
views. Our method most faithfully recovers the body wall, mediastinal
boundary, and fine pulmonary vessels. In comparison, DIF-Net and
DIF-Gaussian exhibit residual streaking and blurring, ILV loses
low-contrast vascular detail, and FDK is dominated by undersampling
artifacts. The neural-field optimizers produce overly smooth volumes
that obscure soft-tissue structure, whereas our reconstruction preserves
both contrast and sharp anatomical boundaries.

\begin{figure*}[!t]
\centering
\includegraphics[width=1.0\linewidth,keepaspectratio,alt={Qualitative comparison at 20 views (same case and layout as Fig.\textasciitilde). Top: DIF-Net, DIF-Gaussian, ILV, ours, ground truth. Bottom: FDK, R\ensuremath{^2}-Gaussian, IntraTomo, NAF, SAX-NeRF, ours, ground truth.}]{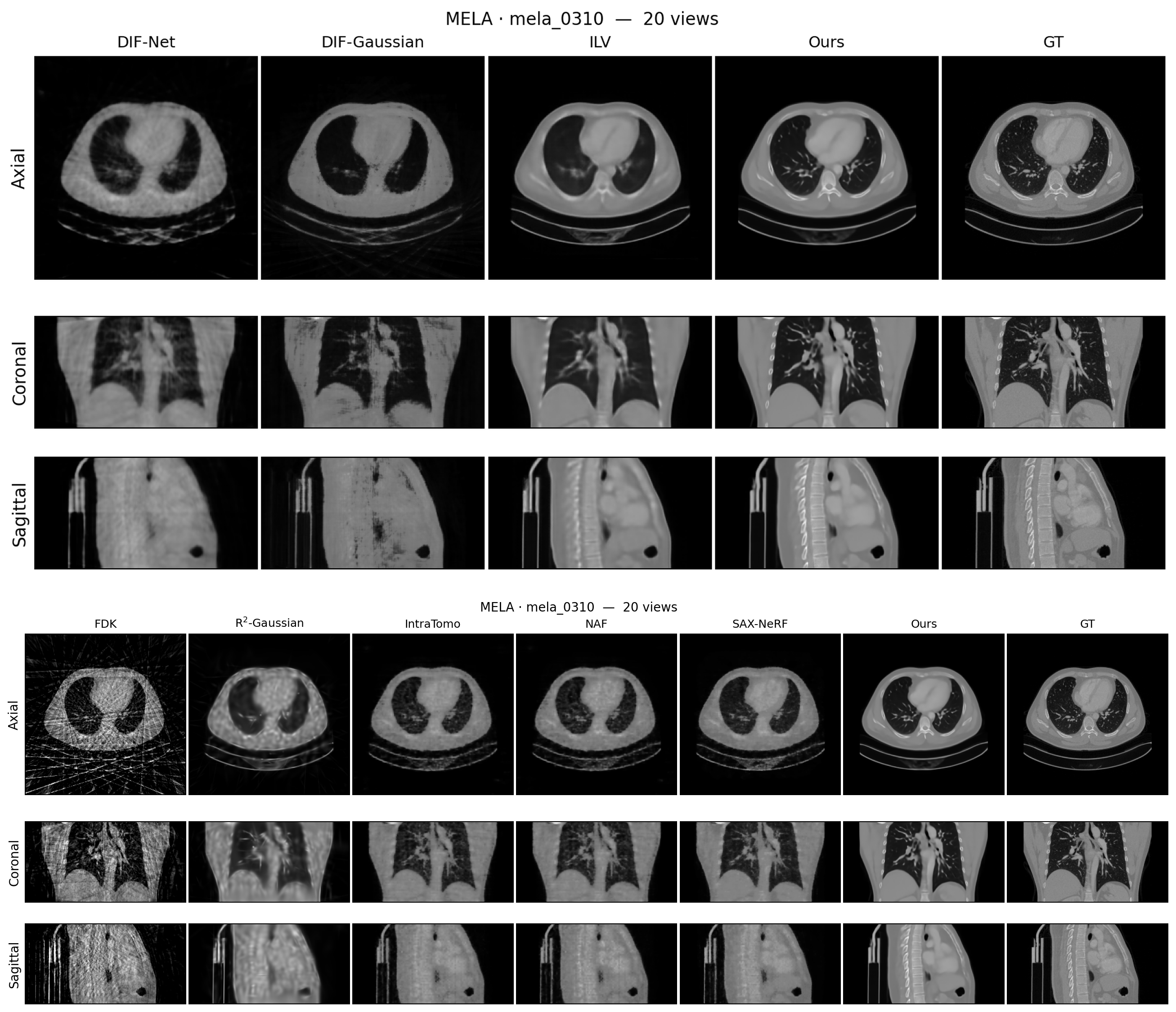}
\caption{\textbf{Qualitative comparison at 20 views} (same case and
layout as Fig.~\ref{fig:qual}). \emph{Top:} DIF-Net,
DIF-Gaussian, ILV, ours, ground truth. \emph{Bottom:} FDK, R\ensuremath{^2}-Gaussian,
IntraTomo, NAF, SAX-NeRF, ours, ground truth.}\label{fig:qual20}
\end{figure*}

\begin{figure*}[!t]
\centering
\includegraphics[width=1.0\linewidth,keepaspectratio,alt={Qualitative comparison at 50 views (same case and layout as Fig.\textasciitilde). Top: DIF-Net, DIF-Gaussian, ours, ground truth. Bottom: FDK, R\ensuremath{^2}-Gaussian, IntraTomo, NAF, SAX-NeRF, ours, ground truth. ILV is not shown: it is trained per view count and no 50-view model was trained.}]{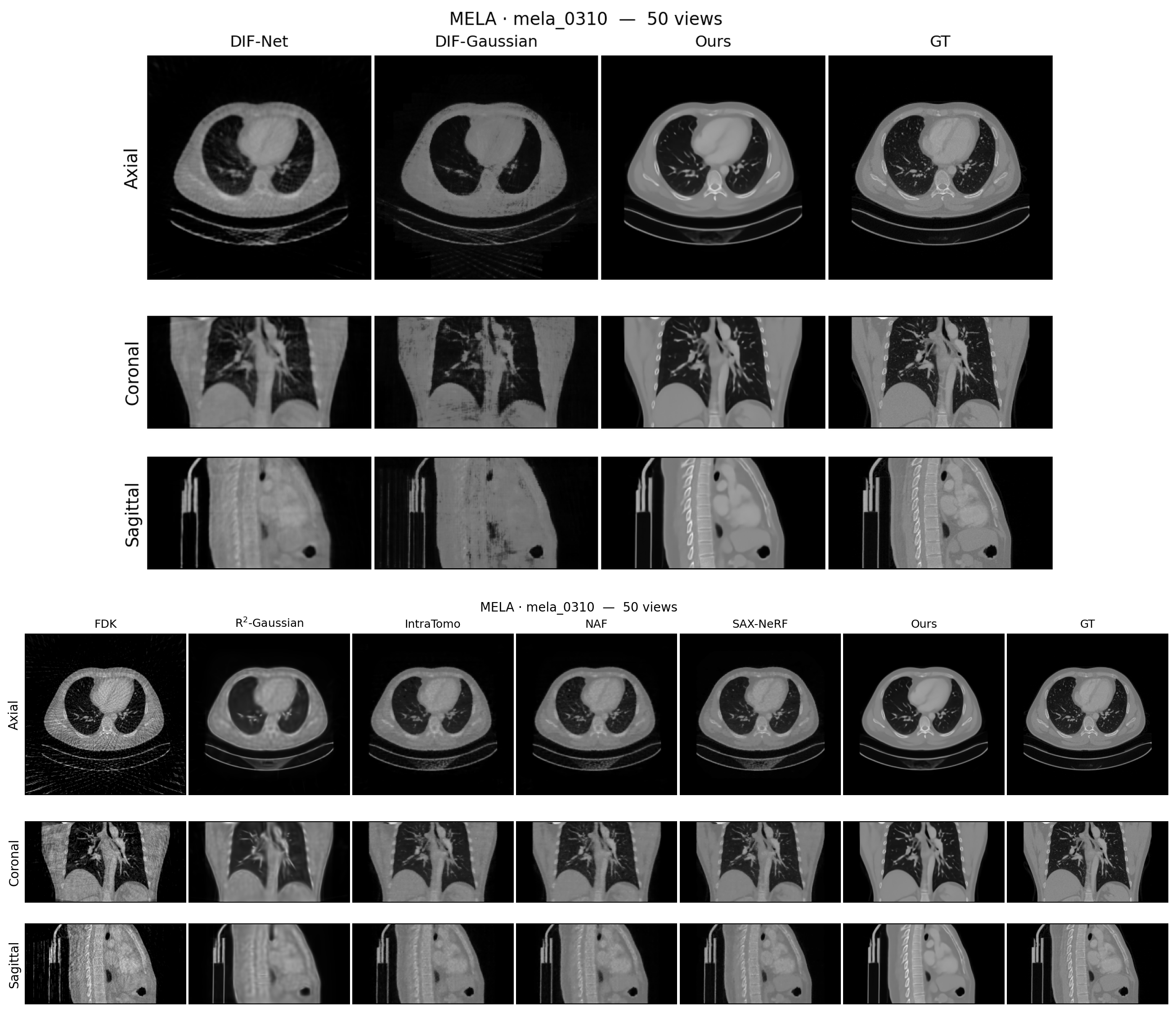}
\caption{\textbf{Qualitative comparison at 50 views} (same case and
layout as Fig.~\ref{fig:qual}). \emph{Top:} DIF-Net,
DIF-Gaussian, ours, ground truth. \emph{Bottom:} FDK, R\ensuremath{^2}-Gaussian,
IntraTomo, NAF, SAX-NeRF, ours, ground truth. ILV is not shown: it is
trained per view count and no \(50\)-view model was
trained.}\label{fig:qual50}
\end{figure*}

\begin{figure*}[!t]
\centering
\includegraphics[width=1.0\linewidth,keepaspectratio,alt={Qualitative comparison at 100 views, measurement-driven methods and ours (same case as Fig.\textasciitilde): FDK, R\ensuremath{^2}-Gaussian, IntraTomo, NAF, ours, ground truth. No feed-forward baseline supports 100 views.}]{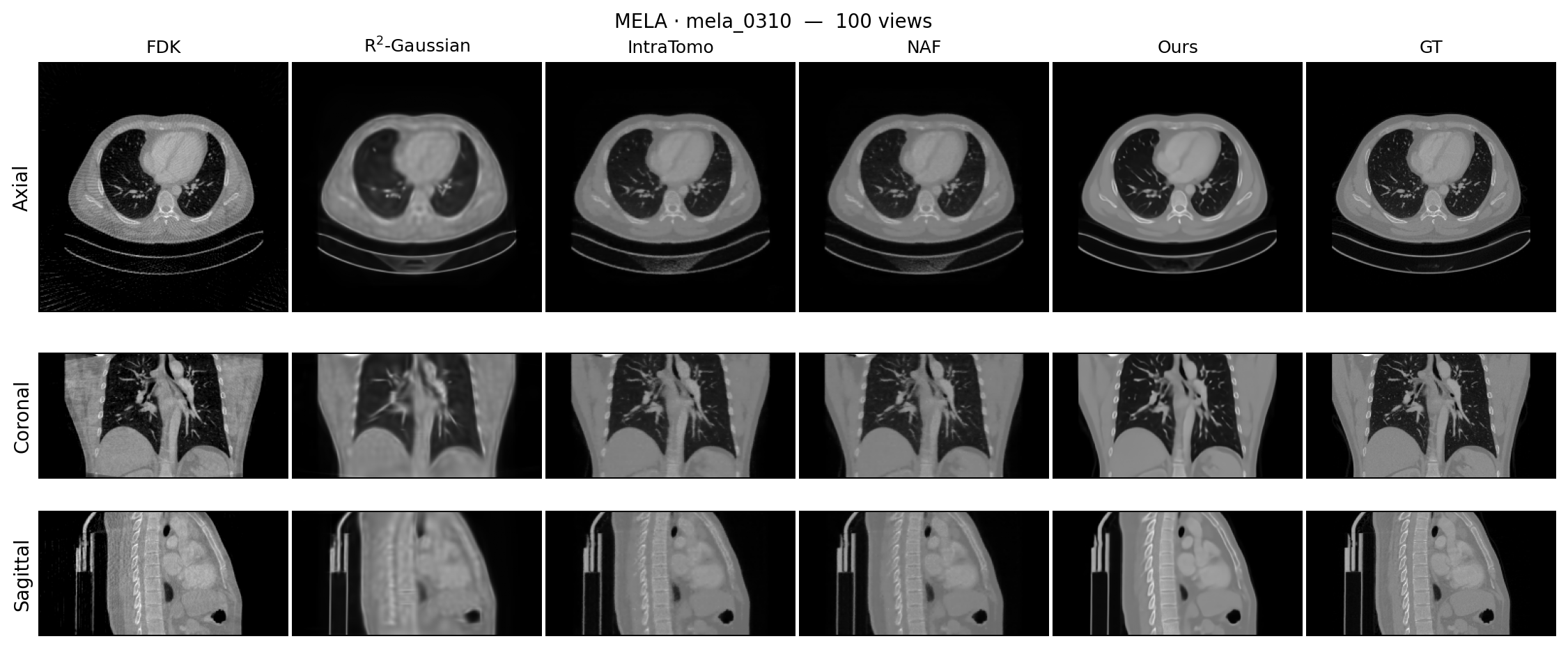}
\caption{\textbf{Qualitative comparison at 100 views},
measurement-driven methods and ours (same case as
Fig.~\ref{fig:qual}): FDK, R\ensuremath{^2}-Gaussian, IntraTomo, NAF,
ours, ground truth. No feed-forward baseline supports \(100\)
views.}\label{fig:qual100}
\end{figure*}

As the view count rises the methods converge, but the ordering is
unchanged: Figs.~\ref{fig:qual20}--\ref{fig:qual100}
repeat the comparison at \(20\), \(50\), and \(100\) views for the same
case. Streaking in the measurement-driven reconstructions recedes
steadily, while the feed-forward baselines retain the blurring of
low-contrast structure that is already visible at \(10\) views. Panels
are omitted where a method does not support the view count.

\subsection{Reconstruction across clinical acquisition
geometries}\label{reconstruction-across-clinical-acquisition-geometries}

The same framework accommodates other clinically motivated acquisition
protocols. Across geometries, the architecture, training procedure, and
loss remain unchanged; only the projection angles selected by the data
loader differ. We train one model per trajectory, and within each, a
single set of weights covers all evaluated view counts. In clinical
practice the trajectory is fixed by the machine and the imaging
protocol, so per-trajectory training matches deployment; a single model
trained jointly across trajectories is left for future work. We apply
this framework to four additional clinically motivated settings
(Fig.~\ref{fig:acq_geom}): two orthogonal views, four
fixed gantry angles (\(0^\circ/45^\circ/90^\circ/135^\circ\)), and
\(90^\circ\) and \(120^\circ\) short arcs. These experiments demonstrate
feasibility across acquisition geometries rather than serve as
head-to-head benchmarks.

\begin{table}[!t]
\scriptsize\setlength{\tabcolsep}{3pt}
\noindent \textbf{Table III. Quality across acquisition geometries (same
architecture and recipe, one model per geometry; reserved test).}
\par\smallskip
\centering
\begin{tabular}{@{}llll@{}}
\toprule\noalign{}
Acquisition & Views & PSNR (dB) & SSIM \\
\midrule\noalign{}
Two orthogonal views & 2 & 25.43 & 0.833 \\
Four fixed views (\(0/45/90/135\ensuremath{^\circ}\)) & 4 & 28.23 & 0.879 \\
\(90\ensuremath{^\circ}\) short arc & 15/30/45 & 27.84/28.36/28.71 & 0.882/0.893/0.900 \\
\(120\ensuremath{^\circ}\) short arc & 15/30/45 & 29.12/29.91/30.05 & 0.896/0.911/0.913 \\
\bottomrule\noalign{}
\end{tabular}
\end{table}

Without any architectural changes, the same framework achieves
\(25.4\)--\(30.1\)~dB PSNR and \(0.83\)--\(0.91\) SSIM
across all acquisition settings. The results also show that angular
coverage matters more than view count alone: four views distributed over
\(135^\circ\) outperform \(15\) views confined to a \(90^\circ\) arc,
while the \(120^\circ\) arc consistently outperforms the \(90^\circ\)
arc at matched view counts. This trend is consistent with broader
angular coverage reducing the missing-wedge ambiguity that cannot be
resolved by increasing the number of views within a limited arc.

\subsection{Real-scan test: real clinical
projections}\label{real-scan-test-real-clinical-projections}

We evaluate the DRR-trained model and every comparator on the \(15\)
clinical Halcyon HyperSight scans described earlier, drawing sparse-view
subsets per scan. At \(10\)--\(100\) views, each method sees only
\(2.5\)--\(30\%\) of the \(335\)--\(407\) projections from which the
clinical reference was reconstructed.

\begin{table}[!t]
\scriptsize\setlength{\tabcolsep}{3pt}
\noindent \textbf{Table IV. Real-data quality (PSNR dB / SSIM; 15 Halcyon
HyperSight scans).}
\par\smallskip
\centering
\begin{tabular}{@{}llllll@{}}
\toprule\noalign{}
Method
 & 
Type
 & 
10 views
 & 
20 views
 & 
50 views
 & 
100 views
 \\
\midrule\noalign{}
\textbf{Ours} & FF & \textbf{25.75 / 0.780} & \textbf{26.44 / 0.806} &
\textbf{26.94 / 0.829} & \textbf{27.24 / 0.841} \\
ILV & FF & 25.19 / 0.725 & 25.62 / 0.751 & OOM & OOM \\
DIF-Net & FF & 23.57 / 0.666 & 23.82 / 0.679 & 24.72 / 0.696 & n/a \\
DIF-Gaussian & FF & 24.28 / 0.665 & 23.19 / 0.557 & 22.53 / 0.506 &
n/a \\
FDK & MD & 15.70 / 0.190 & 17.78 / 0.259 & 20.70 / 0.431 & 22.32 /
0.582 \\
R\ensuremath{^2}-Gaussian & MD & 22.58 / 0.427 & 24.36 / 0.504 & 25.43 / 0.611 & 25.64
/ 0.651 \\
IntraTomo & MD & 24.20 / 0.705 & 25.28 / 0.756 & 26.03 / 0.790 & 26.59 /
0.815 \\
NAF & MD & 24.32 / 0.706 & 25.45 / 0.758 & 26.45 / 0.805 & 26.91 /
0.827 \\
SAX-NeRF & MD & 23.62 / 0.678 & 24.68 / 0.720 & 25.45 / 0.771 & 25.77 /
0.799 \\
\bottomrule\noalign{}
\end{tabular}
\end{table}

Our model achieves the highest PSNR and SSIM at every view count,
although its margins are narrower than on synthetic data: it leads ILV
by \(0.56\)~dB at \(10\) views and NAF by
\(0.33\)~dB at \(100\) views. The relative performance
shift reflects the synthetic-to-real domain gap. Feed-forward methods
trained on DRRs are affected by discrepancies between simulated and
measured projections, whereas measurement-driven methods fit each
measured scan directly and are therefore less sensitive to
training-domain mismatch. For example, our \(10\)-view PSNR changes from
\(29.93\) to \(25.75\)~dB, while that of NAF changes from
\(21.76\) to \(24.32\)~dB. These cross-domain changes
should therefore be interpreted as differences in sensitivity to domain
shift, rather than as a direct comparison of solver quality.

\section{Ablation Studies}\label{ablation-studies}\label{sec:ablation}

Three controlled studies isolate the contributions of model design
(Study 1), training-data scale (Study 2), and the trade-off of using a
single variable-view model instead of view-count-specific specialists
(Study 3). Each study varies only the factor of interest, with its
training and evaluation settings reported alongside the results.

\subsection{Study 1: leave-one-out architecture and loss
factors}\label{study-1-leave-one-out-architecture-and-loss-factors}

Every arm shares an identical encoder, schedule, and stratified
\(1{,}000\)-case training subset (\(100{,}000\) optimizer steps), and is
still evaluated on the full reserved test under the routine
patient-setup protocol at \(N\in\{10,20,50\}\). Each arm differs from
the baseline (A0) by exactly one factor (Table~V), with
one training seed per arm. PSNR is the primary endpoint for every arm;
SSIM and, where reported, soft-tissue error are secondary.

\begin{table}[!t]
\scriptsize\setlength{\tabcolsep}{3pt}
\noindent \textbf{Table V. Study-1 leave-one-out arms (PSNR dB and SSIM, reserved
test).}
\par\smallskip
\centering
\begin{tabular}{@{}llll@{}}
\toprule\noalign{}
Arm
 & 
Change vs.~baseline (A0)
 & 
PSNR \(N{=}10/20/50\)
 & 
SSIM \(N{=}10/20/50\)
 \\
\midrule\noalign{}
A0 & baseline: render off, gradient on, FDK on & 27.90/29.52/31.05 &
0.8646/0.8908/0.9161 \\
A1 & + render (DRR consistency) loss & 27.83/29.43/30.88 &
0.8704/0.8961/0.9199 \\
A2 & \ensuremath{-} multi-scale gradient (MAGA) loss & 27.55/29.17/30.73 &
0.8664/0.8926/0.9184 \\
A3 & Zero out FDK reference (branch kept) & 27.60/28.99/30.10 &
0.8653/0.8883/0.9085 \\
\bottomrule\noalign{}
\end{tabular}
\end{table}

\textbf{A1 --- projection-consistency render loss.} A rendering loss,
which is common practice in feed-forward reconstruction training,
encourages measurement agreement by reprojecting the predicted volume:
\(\mathcal{L}_{\mathrm{render}}=\lambda_{r,1}\lVert\widetilde{\mathcal{P}(\hat{\mathbf V})}-\widetilde{\mathbf p}\rVert_1+\lambda_{r,s}[1-\operatorname{SSIM}(\widetilde{\mathcal{P}(\hat{\mathbf V})},\widetilde{\mathbf p})]\),
where \(\hat{\mathbf V}\) is the predicted volume, \(\mathbf p\) denotes
the input projections, \(\mathcal{P}\) is a differentiable ray-tracing
operator, and \(\widetilde{\cdot}\) denotes per-image min--max
normalization. We set \(\lambda_{r,1}=\lambda_{r,s}=1\) in A1 and
\(\lambda_{r,1}=\lambda_{r,s}=0\) in A0.

Adding this loss to direct volume supervision produced negligible,
metric-dependent changes (PSNR \(-0.11\)~dB; SSIM
\(+0.004\)--\(0.006\)) while increasing per-step training time by
\(2.1\times\)--\(3.1\times\) across GPUs. We attribute the limited
benefit to the stronger spatial supervision provided by the volume
target. Like an algebraic update \cite{andersen1984sart}, a projection
residual propagates its gradient along an entire ray, whereas volume
supervision constrains each voxel directly. We therefore omit the
rendering loss, substantially reducing training cost without a
meaningful loss of reconstruction quality.

\textbf{A2 --- multi-scale gradient (MAGA) loss.} Voxel-wise objectives
can favor over-smoothed reconstructions \cite{lin2025care}, whereas
MAGA directly penalizes boundary and edge errors. Ablating MAGA reduces
PSNR by approximately \(0.3\)~dB across all view counts.
We also evaluate soft-tissue mean absolute error (ST-MAE), defined as HU
MAE within a ground-truth mask spanning
\([-160,240]\)~HU. Removing MAGA increases ST-MAE by
\(1.4\)~HU at \(N=10\) and \(1.3\)~HU at
\(N=50\), confirming poorer soft-tissue fidelity. SSIM marginally favors
the ablated model by \(0.002\) at each view count, consistent with its
known smoothing bias. We therefore retain MAGA based on the PSNR and
targeted soft-tissue improvements rather than unanimous agreement across
metrics.

\textbf{A3 --- in-frame FDK reference.} The FDK branch provides the
model's measurement-derived reference and produces the largest ablation
effect. To isolate its information from its capacity, we zero the FDK
input while retaining the full encoder and fusion branch. Removing the
reference reduces PSNR by \(0.29\), \(0.53\), and
\(0.95\)~dB at \(N=10\), \(20\), and \(50\),
respectively, with the benefit increasing as more projections become
available. SSIM shows the same increasing trend, although it is
essentially unchanged at \(10\) views. This pattern reflects the
improving fidelity of FDK with denser sampling, which allows the model
to exploit an increasingly reliable analytic reconstruction. At \(10\)
views, the standalone FDK image is dominated by severe streak artifacts
and achieves only \(15.34\)~dB (Table~II);
nevertheless, its inclusion yields a measurable \(0.29\)-dB PSNR gain.
EUReCA can therefore extract useful measurement-aligned cues from FDK
even when its image quality is inadequate for standalone use.

\subsection{Study 2: training-set-size
scaling}\label{study-2-training-set-size-scaling}

Feed-forward CBCT methods are increasingly trained on larger datasets,
but the contribution of data scale relative to architectural development
remains unclear. Because published models use substantially different
training-set sizes, cross-study performance is not directly comparable,
and improvements cannot be attributed to architecture alone. We isolate
the effect of data volume using nested, source-stratified subsets
(\(250 \subset 500 \subset 1000 \subset 2000 \subset 4000 \subset 9804\))
while holding the architecture and training recipe fixed. Equal
iteration budgets would over-train smaller subsets while leaving larger
ones under-converged; instead, each model is trained to its validation
plateau, and its best checkpoint is evaluated on the same reserved test
set. These six experiments required approximately \(2{,}300\) GPU-hours
across A40, L40S, A100, H100, and H200 GPUs.

\begin{figure}[!t]
\centering
\includegraphics[width=1.0\linewidth,keepaspectratio,alt={Convergence-matched data-size scaling. PSNR (a) and SSIM (b) versus number of unique training cases (log scale) at N=10/20/50 views, best-validation checkpoints.}]{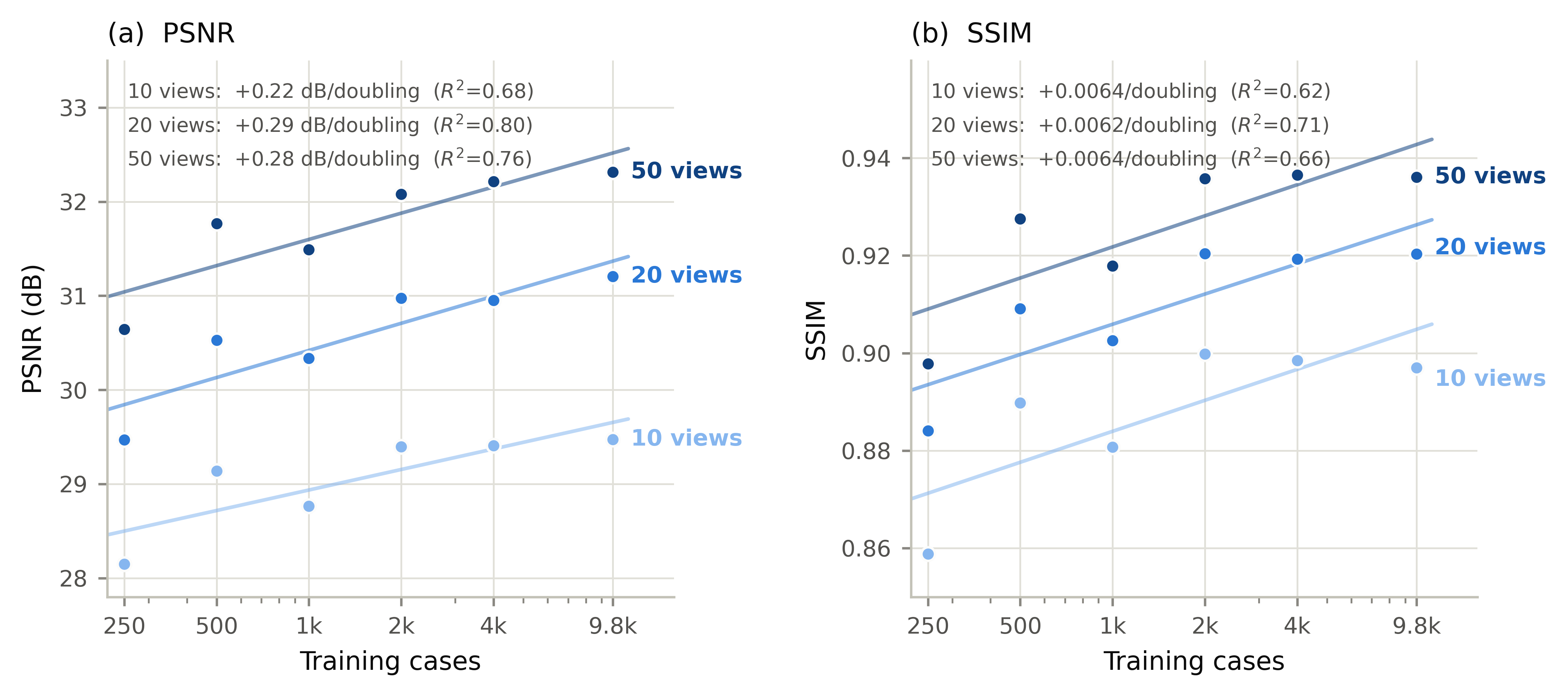}
\caption{\textbf{Convergence-matched data-size scaling.} PSNR (a) and
SSIM (b) versus number of unique training cases (log scale) at
\(N=10/20/50\) views, best-validation
checkpoints.}\label{fig:data_scaling}
\end{figure}

Performance improves substantially throughout the tested range
(Fig.~\ref{fig:data_scaling}). Expanding the training
set from \(250\) to \(9{,}804\) cases increases PSNR by \(1.32\),
\(1.74\), and \(1.67\)~dB at \(10\), \(20\), and \(50\)
views, respectively. Log-linear fits yield
\(0.22\)--\(0.29\)~dB per doubling
(\(R^2=0.68\)--\(0.80\)), equivalent to approximately
\(0.9\)~dB per decade, with no PSNR saturation over the
observed range. SSIM begins to flatten earlier and is therefore less
informative at larger scales. Because each setting uses one seed,
endpoint gains are more reliable than differences between adjacent
subsets. Overall, the sustained PSNR improvement indicates that training
data remain a limiting factor and that the model has not exhausted the
benefits of scale. Larger training corpora therefore offer a clear
opportunity for further performance gains.

\subsection{Study 3: fixed-count specialists vs.~the single combined
model}\label{study-3-fixed-count-specialists-vs.-the-single-combined-model}

To quantify the cost of variable-view training, we compare the combined
model with four specialists trained at individual view counts using an
otherwise identical recipe. Differences are computed per case on the
same \(1{,}226\)-volume test set.

\begin{table}[!t]
\footnotesize
\noindent \textbf{Table VI. Specialists vs.~the combined model (per-case paired
differences, PSNR dB, \(n=1226\); \(\Delta=\) specialist \(-\)
combined). }
\par\smallskip
\centering
\begin{tabular}{@{}lll@{}}
\toprule\noalign{}
Test N & \(\Delta\) PSNR (dB) & 95\% CI \\
\midrule\noalign{}
10 & \(+0.08\) & \([-0.00, +0.17]\) n.s. \\
20 & \(+0.30\) & \([+0.19, +0.41]\) \\
50 & \(+0.25\) & \([+0.12, +0.39]\) \\
100 & \(+0.75\) & \([+0.59, +0.91]\) \\
\bottomrule\noalign{}
\end{tabular}
\end{table}

specialists provide a modest on-count advantage that generally increases
with view count, although the difference at \(10\) views is not
significant. This pattern likely reflects capacity specialization: each
specialist learns the artifact and noise distribution of one sampling
regime, whereas log-uniform training across \(8\)--\(100\) views
distributes capacity across regimes and provides less exposure to the
dense end. Nevertheless, the combined model outperforms the
corresponding specialist on \(28\)--\(41\%\) of individual cases.

The combined model also avoids maintaining and deploying multiple
count-specific networks. Training four specialists requires
approximately four times the compute while supporting only four view
counts, whereas the combined model covers the full range and remains
superior to all external baselines in the main comparison. A
count-balanced sampling curriculum and view-count-specific fine-tuning
may narrow the remaining specialization gap and are left for future
work.

The choice therefore depends on the clinical operating regime. For a
service that acquires a fixed, dense protocol of about \(50\)--\(100\)
views, a specialist (or a count-balanced curriculum) recovers
\(0.25\)--\(0.75\)~dB. For sparse or mixed protocols,
where the view count varies between scans or stays near \(10\)--\(20\)
views, the combined model is preferable: its deficit is small or not
significant, and one network serves every count.

\section{Discussion and Conclusion}\label{discussion-and-conclusion}

Fast reconstruction from few projections has implications beyond imaging
dose reduction. By avoiding a full rotation and per-scan optimization,
EUReCA enables near-real-time volumetric reconstruction across sparse
acquisition settings, which could support repeated intrafraction imaging
for motion analysis, target tracking, or treatment adaptation; none of
these applications was evaluated here, and each would need its own
validation. Its feed-forward design shifts computation to training while
combining learned anatomical priors with analytic estimates derived
directly from the projections. In this way, EUReCA complements classical
reconstruction across diverse clinical acquisition settings. The
complementarity can also run in the other direction: where time allows,
an EUReCA reconstruction could initialize a per-scan optimizer instead
of the uninformative volume such methods start from, leaving the
iterations to enforce projection consistency rather than recover gross
structure. This may improve the final result and supply the explicit
data consistency a feed-forward prediction lacks.

In sum, EUReCA is a geometry-aware, measurement-anchored model for
diverse incomplete CBCT acquisitions. On both simulated data and real
patient data, our method outperforms all applicable feed-forward and
measurement-driven methods across sparse and well-sampled regimes, while
reconstructing in \(0.3\)--\(1.3\)~s. By reproducing a
clinical acquisition geometry and testing on measured data, this work
narrows the gap between simulation and practice. Training with measured
projections is needed to capture realistic noise and system
characteristics, potentially through projection-domain self-supervision
or high-quality full-scan references when paired ground truth is
unavailable. Evaluation should also extend across institutions,
machines, and protocols, and a single model trained jointly across
trajectories would remove the need for one model per trajectory. Pooling
views into angular sectors is itself a lossy reduction: views that share
a sector are collapsed into one feature per voxel, which can cost fine
detail wherever the acquisition is dense enough for a sector to hold
several of them, and the effect grows with the view count. The partition
is a design parameter rather than a property of the method, and the
three scales need not share it; we are exploring finer partitions at the
finer scales, where small angular differences carry the parallax that
resolves detail, while keeping a coarser partition where global context
matters. We studied angular undersampling only: projection noise at
reduced exposure per view was not varied, so robustness to the number of
views does not by itself establish robustness to low-dose acquisitions.
Quantitative Hounsfield unit accuracy remains unverified because the
normalized FDK input provides structural rather than calibrated
intensity information. Calibrated uncertainty and out-of-distribution
detection are also needed to flag weakly supported regions that
pixel-wise metrics may overlook \cite{lin2025care}, with EUReCA's
coverage signal providing a potential basis. More broadly, adapting the
balance between learned priors and measurement evidence across
acquisition regimes is a step toward a general-purpose clinical
reconstruction model.

\bibliographystyle{IEEEtran}
\bibliography{refs}

\end{document}